\documentclass[a4paper,11pt]{article}
\usepackage{amsmath,amsfonts,amssymb,amsthm,amstext,amscd}
\usepackage{latexsym}
\usepackage[hidelinks=true]{hyperref}
\usepackage{graphicx}
\usepackage{caption}
\usepackage{comment}
\usepackage{graphicx}
\usepackage{cite}
\usepackage{color}
\usepackage{setspace}
\usepackage{float}
\usepackage[mathscr]{eucal}
\usepackage{xcolor}
\usepackage{titletoc}
\dottedcontents{section}[1.2em]{}{1em}{1pc}

\newcommand{\mrd}{\mathrm{d}}

\definecolor{darkred}{rgb}{0.9,0.05,0.05}
\definecolor{darkblue}{rgb}{0.05,0.05,0.6}
\definecolor{darkgreen}{rgb}{0.05,0.6,0.05}
\definecolor{brightgreen}{rgb}{0.1,0.9,0.1}

\renewcommand*{\eqref}[1]{%
  \begingroup
    \hypersetup{
      linkcolor=darkblue,
      linkbordercolor=darkblue,
    }%
    \textcolor{darkblue}{(\ref{#1})}%
  \endgroup 
}

\hypersetup{linkcolor=red,citecolor=brightgreen,urlcolor=black,colorlinks=true}

\numberwithin{equation}{section}

\begin{document}

\setlength{\skip\footins}{0.8cm}

\begin{titlepage}

\begin{flushright}\vspace*{-1cm}
{\small
%{\tt arXiv:yymm.nnnn} \\
IPM/P-2016/020 \\
June 6, 2016}\end{flushright}
\vspace{0.5cm}

\begin{center}
\renewcommand{\baselinestretch}{1.5}  %Line spacing
\setstretch{1.5}

{\fontsize{17pt}{20pt}\bf{Revisiting Conserved Charges in Higher Curvature Gravitational Theories}}
 
\vspace{9mm}
\renewcommand{\baselinestretch}{1}  %Line spacing
\setstretch{1}

\centerline{\large{M. Ghodrati$^\dagger$\footnote{e-mail: ghodrati@umich.edu}, K. Hajian$^\ddagger$\footnote{e-mail: kamalhajian@ipm.ir}, M. R. Setare$^\S$\footnote{e-mail: rezakord@ipm.ir} }}

\vspace{3mm}
\normalsize
\textit{$^\dagger$Michigan Center for Theoretical Physics, Randall Laboratory of Physics,\\ University of Michigan, Ann Arbor, MI 48109-1040, USA}\\
 \textit{$^\ddagger$School of Physics, Institute for Research in Fundamental
Sciences (IPM), \\P.O. Box 19395-5531, Tehran, Iran}\\
\textit{$^\S$Department of Science, Campus of Bijar, University of Kurdistan,
Bijar , Iran}
\vspace{5mm}

%%%%%%%%%%%%%%%%%%%%%%%%%%%%%%%%%%%%%%%%%%%%%%%%%%%%%%%%%%%%%%%%%%%%%%%%%%%%%%%%%%%%%%%%%%

%\renewcommand{\baselinestretch}{1.05}  %Line spacing
%\setstretch{1.3}

%%%%%%%%%%%%%%%%%%%%%%%%%%%%%%%%%%%%%%%%%%%%%%%%%%%%%%%%%%%%%%%%%%%%%%%%%%%%%%%%

\begin{abstract}
\noindent
Restricting the covariant gravitational phase spaces to the manifold of parametrized families of solutions, the mass, angular momenta, entropies, and electric charges  can be calculated by a single and simple method. In this method, which has been called ``solution phase space method," conserved charges are unambiguous and regular. Moreover, assuming the generators of the charges to be exact symmetries, entropies and other conserved charges can be calculated on almost arbitrary surfaces, not necessarily horizons or asymptotics. Hence, the first law of thermodynamics would be a local identity relating the exact symmetries to which the mass, angular momentum, electric charge, and entropy are attributed. In this paper, we apply this powerful method to the $f(R)$ gravitational theories accompanied by the terms quadratic in the Riemann and Ricci tensors. Furthermore, conserved charges and the first law of thermodynamics for some of their black hole solutions are exemplified. The examples include warped AdS$_3$, charged static BTZ, and 3-dimensional $z=3$ Lifshitz black holes. 

\end{abstract}
\end{center}
\vspace*{-1.2cm}

\renewcommand{\baselinestretch}{1.05}  %Line spacing
\setstretch{1.5}
\hypersetup{linkcolor=black,citecolor=darkgreen,urlcolor=darkgreen,colorlinks=true} 
\textcolor{black}{\tableofcontents}
\hypersetup{linkcolor=darkred,citecolor=darkgreen,urlcolor=darkgreen,colorlinks=true} 
\end{titlepage}

%%%%%%%%%%%%%%%%%%%%%%%%%%%%%%%%%%%%%%%%%%%%%%%%%%%%%%%%%%%%%%%%%%%%%%%%%%%%%%%%%%%%%%%%%%%
%%%%%%%%%%%%%%%%%%%%%%%%%%%%%%%%%%%%%%%%%%%%%%%%%%%%%%%%%%%%%%%%%%%%%%%%%%%%%%%%%%%%%%%%%%
%\renewcommand{\baselinestretch}{1.05}  %Line spacing
%\setstretch{1.3}

%%%%%%%%%%%%%%%%%%%%%%%%%%%%%%%%%%%%%%%%%%%%%%%%%%%%%%%%%%%%%%%%%%%%%%%%%%%%%%%%
\setlength{\skip\footins}{1cm}

\section{Introduction}\label{sec Intro}
Since the realization of general relativity and Noether's theorems in 1915, there have been numerous attempts to attribute local (and later quasi-local) conserved charges to the symmetries in the presence of gravity. Nowadays, a century after that, the literature on this subject is rich and well established, but still in progress: local conserved charges have not been consistently formulated, while there are different successful formulations for quasi-local conserved charges (see Refs. \cite{Szabados:2004vb,Banados:2016zim} as reviews). Among the different approaches, two main lines of formulation can be distinguished: one is the Hamiltonian formulation which is based on space+time decomposition, and the other one is the Lagrangian formulation which is based on spacetime covariance. Reviewing the timeline of the major progress in these two formulations (which for sure might miss some interesting contributions) can give us an overview, in addition to clarify the motivations of the analysis in this paper. 

Precursor of the Hamiltonian formulation was the introduction of quasi-local charges by Komar in 1959  \cite{Komar:1958wp}. In the Komar's method, quasi-local mass and angular momentum for asymptotic flat solutions could be found by an integration over a codimension-$2$ surface at constant time asymptotics. Soon after, the Hamiltonian formulation of the gravitational theories was elaborated in a series of works in 1959-62 by Arnowitt-Deser-Misner  \cite{Arnowitt:1959ah,Arnowitt:1960es,Arnowitt:1962hi}, known as the ADM formulation. Hence, in addition to introducing a sophisticated formulation for gravitational dynamics, the calculation of the mass and angular momentum at the constant time asymptotics was put on a firm basis (reviewed e.g.  in Ref.  \cite{Gourgoulhon:2007ue}). After that, a similar formulation for the null asymptotics was proposed by Bondi \emph{et al.} in 1962 \cite{Bondi:1962px,Sachs:1962zza}. The Hamiltonian formulation for the asymptotic flat spacetimes reached its mature presentation by Regge-Teitelboim in 1974 \cite{Regge:1974zd}, emphasizing the role of the surface terms in the Hamiltonian. Nonetheless, there was a shortcoming of the formulation, when the asymptotic flatness was relaxed to include asymptotic (anti) de Sitter solutions, mainly because of the appearance of divergent conserved charges. Later progress in this line of formulation has been mainly in the direction of ameliorating this problem (see the review  \cite{Hollands:2005wt}). Transferring to the Hamilton-Jacobi formulation by Brown-York in 1992 \cite{Brown:1992br}, ensued by addition of a surface counterterm to the Lagrangian  \cite{Balasubramanian:1999re}, has been one of them. Reformulation of conserved charges based on  covariantly defined conserved currents, and subtracting the contributions from a reference solution, has been another method proposed by Abbott-Deser in 1982 \cite{Abbott:1981ff}, and this was completed and extended to higher curvature theories by Deser-Tekin   \cite{Deser:2002rt,Deser:2002jk}. This method is known as the ADT method in the literature. Last but not least a contribution has been presented by Kim \emph{et al.} in 2013 \cite{Kim:2013zha}, and it is known as the quasi-local method. It is based on the ADT off-shell conserved current, while two major changes are considered: (1) instead of considering the difference between the solution and a reference solution, a one-parameter integration from the reference solution to the solution under consideration is performed (advocated in Refs.  \cite{Wald:1999wa,Barnich:2003xg}),  (2) the surface of integration is relaxed to be in the interior of the geometry. Hence, it provides a powerful method for calculating conserved charges, specifically for black hole solutions.     

The second line of formulating conserved charges is based on the Lagrangian, which is covariant from the beginning. It was initiated by Ashtekar \emph{et al.} \cite{Ashtekar:1987hia,Ashtekar:1990gc} and Crnkovic-Witten  \cite{Crnkovic:1987at} in 1987, and was consistently formulated in a series of works by Wald \emph{et al.}  \cite{Lee:1990gr,Wald:1993nt,Iyer:1994ys,Wald:1999wa}. In this formulation, which is called covariant phase space formulation (see Refs.  \cite{Hajian:2015eha,Seraj:2016cym,Corichi:2016zac} for reviews), a covariant phase space was built without a space+time decomposition. The phase space manifold was constructed from dynamical fields all over the spacetime, without recruiting their momentum conjugates.  The symplectic form was read from the Lagrangian, which entailed a concrete formulation for conserved charges associated with diffeomorphisms and gauge transformations. Besides, in this progress, the entropy of the non-extermal  \cite{Wald:1993nt,Iyer:1994ys} (and later, of the extremal  \cite{Hajian:2013lna}) black holes was introduced as a conserved charge calculated on black hole horizons. Later in 2002, Barnich-Brandt reformulated the formalism in the language of variational bicomplex, in addition to proposing a version directly from the equation of motion (e.o.m) instead of the Lagrangian  \cite{Barnich:2001jy}. The conceptually strong point (but pragmatically weak point) of the covariant phase space formulation is that the phase space manifold and its tangent space, which are crucial for explicit calculation of conserved charges, are determined by some fall-off conditions. The usual fall-off conditions, although restrict the manifold, do not usually determine it such that calculation of the charges could be performed explicitly. To put the formulation into its full power of calculability, in Ref.  \cite{HS:2015xlp}  the manifold and its tangent space are constructed explicitly and directly from the beginning. The elaboration of the phase space and its tangent space, accompanied by relaxing the calculation of the entropy over horizons, has made the formulation to be a universal tool in the context of conserved charge calculations. This method can be dubbed a \emph{solution phase space method}, because the phase space is constructed by some family of parametrized solutions. Interestingly, the recent independent progresses, the one by Kim \emph{et al.} \cite{Kim:2013zha} in the Hamiltonian formulation, and the solution phase space method \cite{HS:2015xlp} in the Lagrangian formulation, have brought about the two lines of formulations to converge.  

In this paper, we apply the solution phase space method to the higher curvature theories. For clarity, we will focus on $f(R)$ gravity accompanied by quadratic terms in the Riemann and Ricci tensors (see Lagrangian \eqref{Lagrangian scalar}), although the generalization is straightforward. One of the motivations for this work is examining the method for gravitational theories beyond the Einstein-Hilbert gravity. Another motivation is providing detailed materials needed to perform calculations for general enough higher curvature theories. The analysis can be considered in parallel with higher curvature analysis in other methods, specifically the ADT and quasi-local methods studied e.g.  in Refs.  \cite{Deser:2002rt,Deser:2002jk,Nam:2010ub,Devecioglu:2010sf,Alkac:2012bz,Gim:2014nba,Hyun:2014sha,Peng:2014gha,
Fan:2014ala,Bravo-Gaete:2015xea,Setare:2015vea,Setare:2015pva,Setare:2015nla,Myung:2015pua}. In the following sections, first a review on the solution phase space method is presented. Then it is applied to the higher curvature theories. Finally, some interesting examples are provided and compared with the results of other methods. 

\section{Solution phase space method}\label{sec-review}
Solution phase space method (SPSM) is a method for calculating conserved charges in gravitational theories. The SPSM is based on a powerful but not yet fully appreciated covariant formulation of gravitational phase spaces, which we are going to review in the next subsection. It is applicable to the solutions which are parametrized by some parameters $p_j$. Specifically, it is a convenient method for calculating mass, angular momenta, entropies, and electric charges associated with the black hole solutions, although it is not exclusive to them  \cite{Hajian:2015eha,HS:2015xlp}. 

Before delving into the details, it can be helpful to have a look at the big picture and the bottom line: calculation of variation of a conserved charge needs three pieces of information as inputs: (1) the theory in $d$ dimensional spacetime, (2) the solution and some perturbation around it for which charge is calculated, and (3) the symmetry to which the charge is attributed. At the end of the day, integrating a $d-2$-form $\boldsymbol{k}_\eta(\hat\delta\Phi,\hat\Phi)$ over any codimension-$2$ surface yields the variation of the charge. Concerning the three inputs mentioned above, $\boldsymbol{k}$ is unambiguously determined by the theory. Its arguments $\hat\Phi$ and $\hat\delta \Phi$ denote some elaborated solutions and perturbations. $\eta$ carries information as regards the symmetry. If the result would be integrable, then an integration over $\hat\delta \Phi$ produces the finite charge.

\subsection{Covariant phase space formulation} 
Covariant phase space formulation is an appropriate and well-established construction of the gravitational phase spaces \cite{Crnkovic:1987at,Ashtekar:1987hia,Ashtekar:1990gc,Lee:1990gr,Wald:1993nt,Iyer:1994ys}.  To have a self-contained document, we will review the basics of the formulation here, which might have some overlaps with the reviews in Refs. \cite{HS:2015xlp,Hajian:2016kxx}. At the outset, it would be useful recalling some relevant elementary properties of a phase space. Phase space $\mathcal{F}(\mathcal{M},\Omega)$ is a manifold $\mathcal{M}$ equipped with a closed nondegenerate  symplectic form $\Omega$. In order to introduce a physical phase space, one usually begins with a given Lagrangian. For example, dynamics of a particle in one dimension can be described by the Lagrangian $L=\frac{m \dot{\text{q}}^2}{2}-V(\text{q})$.  $\mathcal{M}$ would be built by the position and its momentum conjugate $(\text{q},\text{p})$, equipped with the canonical symplectic $2$-form $\Omega=\delta \text{p}\wedge\delta \text{q}$. A simple way to derive this symplectic form is to vary the Lagrangian
\begin{equation}\label{omega trick 1}
\delta L=(\frac{\partial L}{\partial \text{q}}-\frac{\mrd}{\mrd t}\frac{\partial L}{\partial \dot{\text{q}}})\delta \text{q}+\frac{\mrd}{\mrd t}(\text{p}\,\delta \text{q})\,.
\end{equation}
Then, by recognizing the first term as the e.o.m, and the second term as a total derivative,  $\frac{\mrd}{\mrd t}$ in the latter has to be dropped, and its exterior derivative on the phase space should be taken:
\begin{equation}\label{omega trick 2}
\frac{\mrd}{\mrd t}(\text{p}\,\delta \text{q}) \quad \rightarrow \quad  \text{p}\,\delta \text{q}\quad  \to\quad \delta (\text{p}\,\delta \text{q})= \delta \text{p}\wedge\delta \text{q}\,.
\end{equation}   
Equipped with the $\Omega$, and for a given vector field $v$ on the $\mathcal{M}$, a charge variation $\delta H_v$ can be defined by 
\begin{equation}\label{charge trick}
\delta H_v\equiv v\cdot\Omega \,.
\end{equation}
For instance, in our simple example, choosing $v=\partial_{ \text{q}}$, then $\delta H_v=\partial_{ \text{q}}\cdot \Omega=\delta \text{p}$. Hence, $H_v= \text{p}$, which is the mathematical manifestation of ``momentum is the charge attributed to the translation in space". Notice that in order for a charge to be conserved, one needs extra conditions. In the case of the simple example mentioned above, $\text{p}$ would be conserved if only $V(\text{q})$ would be a constant. 
\paragraph{Covariant phase space:} Let us begin with a given Lagrangian in $d$ dimensional spacetime, with some classical dynamical fields, collectively denoted by $\Phi(x^\mu)$. The fields might include the metric $g_{\alpha\beta}$, some Abelian gauge fields $A^a_\mu$, some scalars $\phi^I$, etc.  It is usual to build the phase space canonically, i.e.  to build the $\mathcal{M}$ from a subset of field configurations $\Phi(\vec{x})$ and their momentum conjugates defined on some privileged time foliation of spacetime. In this construction, solutions to the equation of motion are some \emph{curves} on $\mathcal{M}$ parametrized by the time. Interestingly, in the context of generally covariant gravitational theories, there is a more suitable construction which does not break general covariance by specifying a time foliation. In this construction, $\mathcal{M}$ is composed of dynamical field configurations all over the spacetime $\Phi(x^\mu)$. On the other hand, the field conjugates would not be needed to construct the manifold. As a result, any solution to the equation of motion in the phase space would be a \emph{point} on $\mathcal{M}$, instead of a curve. The tangent space of the manifold is also constituted from  a subset of perturbations $\delta \Phi(x^\mu)$. 

\paragraph{Symplectic structure:} The manifold $\mathcal{M}$ which is constructed is a phase space. The symplectic $2$-form is constructed from the Lagrangian $d$-form $\mathbf{L}$. To this end, using the same method as in Eqs.  \eqref{omega trick 1} and \eqref{omega trick 2}, first the Lee-Wald $(d\!-\!1)$-form $\mathbf{\Theta}$ is picked up from the surface term appearing in the variation of Lagrangian:
\begin{equation}\label{find Theta}
\delta \mathbf{L}=\mathbf{E}_{{\Phi}}\delta \Phi+\mrd \mathbf{\Theta}_{_\text{LW}}(\delta \Phi,\Phi)\,.
\end{equation}
In the equation above,  $\mathbf{E}_{{\Phi}}$ denotes equations of motion for the fields $\Phi$, on which the summation convention should be understood.  $\delta$ and $\mrd$ are exterior derivatives on $\mathcal{M}$ and on spacetime respectively. Then the pre-symplectic form can be defined as  \cite{Lee:1990gr,Wald:1993nt,Iyer:1994ys}
\begin{equation}\label{Omega LW}
\Omega_{_\text{LW}}(\delta_1\Phi,\delta_2\Phi,\Phi)\equiv \int_\Sigma \boldsymbol{\omega}_{_\text{LW}}(\delta_1\Phi,\delta_2\Phi,\Phi)\, 
\end{equation}
where
\begin{equation}\label{omega LW}
\boldsymbol{\omega}_{_\text{LW}}(\delta_1\Phi,\delta_2\Phi,\Phi)=\delta_1\mathbf{\Theta}_{_\text{LW}}(\delta_2\Phi,\Phi)-\delta_2\mathbf{\Theta}_{_\text{LW}}(\delta_1\Phi,\Phi)\,.
\end{equation}      
 The $\Sigma$ is some codimension-1 (Cauchy) surface and $\delta_{1,2}\Phi$ are some members of the tangent space. The $\boldsymbol{\omega}_{_\text{LW}}$, which is a $2$-form over the phase space and a $d-1$-form over the spacetime,  is called pre-symplectic current. By construction, $\Omega$ in Eq. \eqref{Omega LW} is antisymmetric in $\delta_1\Phi\leftrightarrow \delta _2\Phi$, and is a closed form $\delta \Omega=0$. If it would be a non-degenerate form, it could be used to construct a symplectic structure. In this case, one drops the prefix in ``pre-symplectic," and calls it a symplectic form.   
 
\paragraph{Conservation:} Apparently, $\Omega_{_\text{LW}}$ in Eq. \eqref{Omega LW} depends on a non-covariantly chosen surface $\Sigma$. In order to make $\Omega_{_\text{LW}}$ independent of $\Sigma$, which in this context is called ``conservation of symplectic form," one needs $\mrd \boldsymbol{\omega}_{_\text{LW}}=0$. Moreover, the flow of $\boldsymbol{\omega}_{_\text{LW}}$ passing throughout the boundaries $\partial \Sigma$ should vanish. The former is achieved if $\Phi$ and $\delta\Phi$ satisfy the e.o.m and the linearized e.o.m respectively. So, it is standard to restrict  the phase space to the solutions, as we will do in the rest of the paper.  On the other hand, achievement of the latter needs extra conditions, usually some boundary conditions on the perturbations. 
 
\paragraph{Ambiguities:} There are two kinds of ambiguities present in the covariant phase space formulation; one an irrelevant and another one a relevant. The irrelevant one originates from the fact that the formulation is based on Lagrangian formulation, which is itself ambiguous up to a surface term $\mathbf{L}\to \mathbf{L}+\mrd \boldsymbol{\mathcal{K}}$. Nonetheless, although it results to $\mathbf{\Theta}\to \mathbf{\Theta}+\delta \boldsymbol{\mathcal{K}}$, but $\boldsymbol{\omega}$ remains intact because of $\delta ^2 \boldsymbol{\mathcal{K}}=0$. Another ambiguity, which is the relevant one, is an ambiguity originating from the definition of $\mathbf{\Theta}$ in Eq. \eqref{find Theta}; one can add an exact $(d-1)$-form $\mrd\mathbf{Y}(\delta \Phi,\Phi)$ to $\mathbf{\Theta}_{_\text{LW}}(\delta \Phi,\Phi)$, i.e. 
\begin{equation}
\mathbf{\Theta}_{_\text{LW}}(\delta \Phi,\Phi)\to \mathbf{\Theta}(\delta \Phi,\Phi)=\mathbf{\Theta}_{_\text{LW}}(\delta \Phi,\Phi)+\mrd\mathbf{Y}(\delta \Phi,\Phi)\,.
\end{equation}
This ambiguity entails corresponding ambiguities in $\Omega$ defined above, through 
\begin{equation}\label{omega ambiguity}
\boldsymbol{\omega}(\delta_1\Phi,\delta_2\Phi,\Phi)\to \boldsymbol{\omega}(\delta_1\Phi,\delta_2\Phi,\Phi)+ \mrd \big(\delta_2 \mathbf{Y}(\delta_1 \Phi,\Phi)-\delta_1 \mathbf{Y}(\delta_2 \Phi,\Phi)\big)\,.
\end{equation}

\paragraph{Conserved charges:} Let us consider a vector field $\xi=\xi^\mu\partial_\mu$  defined over the spacetime, which generates the diffeomorphism $x^\mu\to x^\mu-\xi^\mu$. In addition, we might have some scalars on the spacetime $\lambda ^a$, generating the gauge transformations $A^a_\mu\to A^a_\mu+\partial_\mu \lambda^a$. We can denote the generator of the combination diffeomorphism+gauge transformations by $\epsilon=\{\xi,\lambda^a\}$ such that $\delta_\epsilon\Phi\equiv \mathscr{L}_\xi\Phi+\delta_{\lambda^a}A^a$. Being equipped with the symplectic form, and motivated by the definition of charge Eq. \eqref{charge trick}, we might be able to associate a conserved charge (interchangeably called Hamiltonian generator) to the $\epsilon$.   To this end, variation of the charge is defined as \cite{Lee:1990gr,Wald:1993nt,Iyer:1994ys,Compere:2015knw}
\begin{align}\label{delta H xi}
\delta H_{\epsilon}(\Phi)&\equiv  \int_\Sigma \big(\delta^{[\Phi]}\mathbf{\Theta}(\delta_\epsilon\Phi,\Phi)-\delta_\epsilon\mathbf{\Theta}(\delta\Phi,\Phi)\big)=\int_{\Sigma}\mrd\boldsymbol{k}_{\epsilon}(\delta\Phi,\Phi)=\oint_{\partial\Sigma}\boldsymbol{k}_{\epsilon}(\delta\Phi,\Phi) \,,
\end{align} 
where the last equation follows from Stokes' theorem. The $\delta^{[\Phi]}$ emphasizes that $\delta$ acts on dynamical fields, not  $\epsilon$.  In the equation above, the integrand in the first integration has been replaced by an exact $(d\!-\!1)$-form $\mrd \boldsymbol{k}_\epsilon$. This is the fundamental theorem of the covariant phase space formalism, which can be proved using the on-shell conditions \cite{Wald:1993nt,Iyer:1994ys}. The $(d\!-\!2)$-form $\boldsymbol{k}_\epsilon$ can be shown to be explicitly (see e.g.  Appendix A in Ref.  \cite{HS:2015xlp} for the detailed derivation)
\begin{equation}\label{k_xi general}
\boldsymbol{k}_\epsilon(\delta\Phi,\Phi)=\delta \mathbf{Q}_\epsilon-\xi \cdot \mathbf{\Theta}(\delta \Phi,\Phi)\,,
\end{equation} 
in which $\mathbf{Q}_\epsilon$ is the \emph{Noether-Wald charge density}  \cite{Wald:1993nt,Iyer:1994ys}, defined by the relation
\begin{equation}\label{Noether Wald}
\mrd \mathbf{Q}_\epsilon\equiv \mathbf{\Theta}(\delta_\epsilon\Phi,\Phi)-\xi \! \cdot \! \mathbf{L}\,.
\end{equation}
Hence, by  Eq.  \eqref{k_xi general},  $\boldsymbol{k}_\epsilon$ can be found for a given theory straightforwardly. Putting it into Eq.  \eqref{delta H xi}, the charge variation $\delta H_{\epsilon}(\Phi)$ can be calculated for an arbitrary generator $\epsilon$. Concerning the conservation, by $\mrd \boldsymbol{\omega}(\delta \Phi,\delta_\epsilon\Phi,\Phi)= \mrd ^2\boldsymbol{k}_\epsilon(\delta \Phi,\Phi)=0$ there would not be any source or sink in $\Sigma$. But vanishing of the flux/leakage  through $\partial\Sigma$ needs to be investigated.

\paragraph{Integrability:}  $\delta H_{\epsilon}(\Phi)$, which is calculated by the last integral in Eq.  \eqref{delta H xi},  might corresponds to the variation of a finite conserved charge $H_{\epsilon}$. In order to investigate this finite conserved charge, integrability over the phase space is needed. This condition is basically $(\delta_1\delta_2-\delta_2\delta_1)H_\epsilon (\Phi)=0$, in which $\Phi$s are any field configuration in the presumed phase space $\mathcal{F}$, and $\delta_{1,2}\Phi$ are any arbitrary chosen member of its tangent space. It follows that this condition can be explained as  \cite{Lee:1990gr,Wald:1999wa,Compere:2015knw}
\begin{equation}\label{integrability cond modified}
\oint_{\partial\Sigma} \Big(\xi\cdot \boldsymbol{\omega}(\delta_1\Phi,\delta_2\Phi,\Phi)+\boldsymbol{k}_{\delta_1\epsilon}(\delta_2\Phi,\Phi) -\boldsymbol{k}_{\delta_2\epsilon}(\delta_1\Phi,\Phi)\Big)=0\,, \qquad \forall \Phi, \delta_{1,2}\Phi\,. 
\end{equation}

\paragraph{Symplectic symmetries:} As far as conserved charges are concerned, conservation of $\delta H_\epsilon$ can be guaranteed if $\epsilon$ is chosen such that
\begin{equation}
\boldsymbol{\omega}(\delta\Phi,\delta_\epsilon\Phi,\Phi)=0
\end{equation}
on-shell. It is because there would not be any flow out of the boundaries locally, and hence globally. The family of $\epsilon$'s with this property, which has been dubbed ``symplectic symmetry generators" \cite{CHSS:2015mza}, can be divided into two sets  \cite{HS:2015xlp}:
\begin{enumerate}
\item \emph{Non-exact symplectic symmetries:} The $\epsilon$ for which $\delta_\epsilon \Phi\neq0$ at least on one point of the phase space. They constitute a closed algebraic structure, and have been proposed to be responsible for generating the phase space of a solution at some given constant thermodynamical variables \cite{CHSS:2015mza,CHSS:2015bca,Compere:2015knw}. Hence, by studying them and the phase space generated via exponentiating them, one might hope to understand the microstates of the system.
\item \emph{Exact symplectic symmetries:} The ones for which $\delta_\epsilon \Phi=0$ all over the phase space. For clarity, let us denote such generators by $\eta=\{\zeta,\lambda^a\}$. They are considered as generators which by their conserved charges, the set of solutions in different thermodynamical variables can be labeled \cite{HS:2015xlp}. The phase space constructed by such field configurations has been called ``solution phase space," \cite{HS:2015xlp} which will be described in a moment.
\end{enumerate}
It has been conjectured that the phase space associated with the geometries without propagating degrees of freedom are direct product of these two families of phase spaces: the statistical phase space $\otimes$ the solution phase space  \cite{HS:2015xlp,Sheikh-Jabbari:2016lzm}.

\subsection{Solution phase space method; conserved charges and the first law(s)} \label{subsec-SPSM}

In the covariant phase space formulation reviewed above, it is standard to identify the phase space manifold by some asymptotic behaviors, usually through requiring some fall-off conditions. Fall-off conditions, although they delimit the phase space manifold,  usually do not determine it completely. ``Solution phase space method" is an alternative method for determining the phase space manifold. Restricting the covariant phase space formulation to some explicitly identified manifolds, empowers the calculability of this formulation. The specification is in three aspects, which will be described immediately:
\begin{enumerate}
\item Identifying the phase space manifold explicitly,
\item Crystallizing the tangent space of the specified manifolds,
\item Concentrating on the exact sympltectic symmetries of the proposed phase space.
\end{enumerate}
Consider a family of (black hole) solutions to a generally covariant gravitational theory. Usually, such a family is identified by some isometries and some parameters $p_j$. The field configuration of such a family can be denoted collectively by $\hat \Phi(x^\mu,p_j)$. The parameters are some arbitrary (but maybe in some restricted domain of) real numbers appearing in the field configuration of the mentioned solutions. The $p_j$ can be reparametrized, but they cannot be removed by coordinate transformations. The manifold $\hat{\mathcal{M}}$ can be chosen to be composed of the members of the family, up to unphysical coordinate/gauge transformations. As an example, the set of all Schwarzschild black holes
\begin{equation}
\mrd s^2=-(1-\frac{2Gm}{r})\mrd t^2+\frac{\mrd r^2}{1-\frac{2Gm}{r}}+r^2 \mrd \theta^2+r^2\sin ^2 \theta \mrd \varphi^2\,,
\end{equation}
parametrized by one free parameter $p_{_1}=m\geq 0$, construct a manifold $\hat{\mathcal{M}}$. 

The symplectic $2$-form $\hat \Omega$ would be simply the Lee-Wald symplectic form \eqref{Omega LW}, which is confined to $\hat{\mathcal{M}}$. Therefore, the $\hat{\mathcal{F}}=(\hat{\mathcal{M}},\hat \Omega)$ would constitute a phase space, the ``solution phase space". Hence, any point of the manifold can be identified by $\hat \Phi(x^\mu,p_j)$. The tangent space of $\hat{\mathcal{M}}$ is spanned (up to infinitesimal pure gauge transformations) by ``parametric variations," which can be found simply by  \cite{HSS:2014twa}
\begin{equation}
\hat{\delta}\Phi= \frac{\partial \hat{\Phi}}{\partial p_j}\delta p_j.
\end{equation} 
These variations, which are infinitesimal difference of two solutions, satisfy linearized equation of motion. As a result, they respect $\mrd\boldsymbol{\omega}_{_\text{LW}}(\hat\delta_1\Phi,\hat\delta_2\Phi,\hat\Phi)=0$. 

In SPSM, the diff+gauge transformations, for which charges are calculated, should be restricted to the symplectic symmetries. However, our main focus would be on the set of exact symmetries. Denoting the generator of the exact symmetries by $\eta=\{\zeta,\lambda^a\}$ such that $\delta_\eta \hat\Phi=0$, the $\zeta$ would be a Killing vector of all points of the phase space $\hat{\mathcal{M}}$. Besides, its action on the gauge fields has to be canceled by the action of $\lambda^a$'s, i.e.  $\mathcal{L}_\zeta A^a_\mu+\partial_\mu \lambda^a=0$. As it was advertised in Sect. \ref{subsec-SPSM}, conservation of $\hat\delta H_\eta$ is guaranteed. This is because of the relation  $\boldsymbol{\omega}_{_\text{LW}}(\hat\delta\Phi,\delta_\eta\hat\Phi,\hat\Phi)=0$, which itself is a result of linearity  of $\boldsymbol{\omega}_{_\text{LW}}$ in $\delta_\eta\hat\Phi=0$. Hence, there would not be any local and, therefore, any global flow of $\boldsymbol{\omega}_{_\text{LW}}$ out of the boundaries $\partial \Sigma$.   

Along with guaranteeing the conservation, focusing on the exact symmetries provides us some other nice features:
\begin{itemize}
\item \textit{Independence of $\hat\delta H_\eta$ from the choice of $\partial \Sigma$:} The relation $\boldsymbol{\omega}_{_\text{LW}}(\hat\delta\Phi,\delta_\eta\hat\Phi,\hat\Phi)=0$ yields an interesting result: $\hat\delta H_\eta$ would be independent of the chosen $\partial\Sigma$. It is because of vanishing of $\boldsymbol{\omega}_{_\text{LW}}$  all over the $\Sigma$, and hence, vanishing of $\boldsymbol{\omega}_{_\text{LW}}$ in the region enclosed between two different integrating surfaces $\partial\Sigma_1$ and $\partial\Sigma_2$. Then, by the Stokes theorem, and taking the result of Eq.  \eqref{delta H xi} into account, the claim is proved.  Explaining  this result in another way, although the integration in calculating $\hat\delta H_\eta$ is over a codimension-2 surface $\partial \Sigma$, but the result would be independent of all coordinates, including the two coordinates which are not integrated over.
\item \textit{Discarding the ambiguity $\mathbf{Y}$:} This is because of $\delta\mathbf{Y}(\delta_\eta\Phi,\Phi)-\delta_\eta\mathbf{Y}(\delta\Phi,\Phi)=0$, which is a result of the linearity  of the left hand side in $\delta_\eta\Phi=0$. Using this identity together with  Eq. \eqref{omega ambiguity} in the definition of charge variations Eq. \eqref{delta H xi}, then there would not be any ambiguity in the calculated conserved charges as far as exact symmetries are considered.
\end{itemize}
 Summarizing the last two paragraphs, the charges associated with exact symmetries are {conserved}, {unambiguous}, and independent of the chosen closed surfaces of integration $\partial \Sigma$.

So far, the SPSM has provided all materials needed to {calculate} conserved charge variations.  The final tasks would be checking integrability over $\hat{\mathcal{M}}$, and (if integrable) performing the integration. The integrability can be assessed by replacing $\Phi$, $\delta\Phi$, and $\epsilon$ in the integrability condition Eq.  \eqref{integrability cond modified} by $\hat\Phi$, $\hat \delta \Phi$, and $\eta$, respectively. If integrable, then the integration over arbitrary path on $\hat{\mathcal{M}}$ connecting a reference field configuration $\hat\Phi(\bar p_j)$ to the solution under consideration $\hat\Phi(p^0_{j})$ yields the final result
\begin{equation}\label{finite H xi}
H_\eta[\hat\Phi(p^0)]-\!H_\eta[\hat\Phi(\bar p)]=\int_{\bar p}^{p^0}\!\hat\delta H_\eta\,.
\end{equation}
$H_\eta[\hat\Phi(\bar p)]$ is the reference point (i.e.  constant of integration) for the $H_\eta$ defined on some specific reference field configuration $\hat\Phi(x^\mu;\bar p_j)$.   

It is worth mentioning that in order to perform the final tasks mentioned above, there is a shortcut: by the conservation$+$independence from $\partial \Sigma$, the $\hat \delta H_\eta$ would only be a function of $p_j$ and $\delta p_j$, not any coordinate of the spacetime. Hence, one can simply check whether it is a total derivative or not. Then, in the case of being a total derivative, the integration can be done by an appropriate choice of a reference field. For instance, if the $\hat{\mathcal{M}}$ is parametrized by two parameters $\{p_{_1},p_{_2}\}$, and one has found $\hat \delta H_\eta=p_{_1}\delta p_{_2}+p_{_2}\delta p_{_1}$, then it is a total derivative $\hat \delta H_\eta=\hat \delta (p_{_1} p_{_2})$. So, the integrated charge would be simply $H_\eta=p_{_1}p_{_2}+\text{const.}$, where the constant would be fixed by the choice of a reference field configuration. 

Before moving on to the next sections, which will provide us explicit examples, it can be useful recalling some remarks in the context of SPSM. 
\begin{itemize}
\item[$\circ$] Not all exact symmetries have integrable conserved charges. Hence, integrability puts constraint on the choice of exact symmetries to which mass, angular momenta etc. are attributed.    
\item[$\circ$] Electric charge associated to the gauge field $A^a$, denoted by $Q_a$, is the conserved charge of the global gauge transformation $\eta_{_{Q_a}}=\{0,1^a\}$ in which $1^a$ means $\lambda^a=1$ and $\lambda^b=0$ for $b\neq a$. 
\item[$\circ$] Similar to the electric charge, mass and angular momenta are conserved charges which are attributed to the geometry as a whole. For the stationary solutions with some number of axial $U(1)$ isometries (labeled by $i$), one can choose the coordinates such that the corresponding Killing vectors would be $\partial_t$ and $\partial_{\varphi^i}$, respectively. Then, up to a conventional normalization, the exact symmetries to which the mass $M$ and angular momenta $J_i$ are attributed would be $\eta_{_M}=\{\partial_t+\Omega^i_{_\infty}\partial_{\varphi^i},-\Phi^a_{_\infty}\}$ and $\eta_{_{J_i}}=\{-\partial_{\varphi^i},0\}$. $\Omega^i_{_\infty}$ and $\Phi^a_{_\infty}$ are asymptotic angular velocities and electric potentials, which are usually adopted to be zero.
\item[$\circ$] In contrast with the charges mentioned above, to each one of the horizons in a geometry, one can associate an entropy. Hence, there might be more than one entropy in a single geometry, e.g.  entropy of inner, outer or cosmological horizons. Entropies are considered to be conserved charges for the exact symmetries $\eta_{_\mathrm{H}}=\frac{2\pi}{\kappa_{_\mathrm{H}}}\{\zeta_{_\mathrm{H}},-\Phi^a_{_\mathrm{H}}\}$, in which $\kappa_{_\mathrm{H}}$, $\zeta_{_\mathrm{H}}$, $\Phi^a_{_\mathrm{H}}$ are surface gravity,  Killing vector, and electric potential of the horizon, respectively. Notice that the $\zeta_{_\mathrm{H}}$ should be accompanied by the rigid gauge transformations $\lambda^a=-\Phi^a_{_\mathrm{H}}$, and be normalized by the surface gravity, in order to have an integrable charge.
\item[$\circ$] It is worth emphasizing again that assuming the generators of mass, angular momenta, electric charges and entropies to be exact symmetries, these charges can be calculated by integrations over almost arbitrary $\partial \Sigma$, and not necessarily the horizons or asymptotics. In this respect, the entropies are on equal footing with other conserved charges.
\end{itemize} 
\paragraph{First law(s) of thermodynamics:}
To each one of the horizons denoted by ``H", an entropy $S_{_\mathrm{H}}$, temperature $T_{_\mathrm{H}}=\frac{\kappa_{_\mathrm{H}}}{2\pi}$, and some chemical potentials $\Omega_{_\mathrm{H}}^i , \Phi^a_{_\mathrm{H}}$ etc. can be attributed  \cite{Bekenstein:1973ft,Bardeen:1973gd,Hawking:1976rt}. The first law of thermodynamics for the chosen horizon relates  $\delta S_{_\mathrm{H}}$ to the variations of other conserved charges attributed to the whole geometry. In the SPSM, derivation of the first law(s) is very simple, and originates from a \emph{local} identity;  $\eta_{_\mathrm{H}}$ is a linear combination of the generators of mass, angular momenta, and electric charges. From this identity, the first law follows by the linearity of the generic charge variations $\delta H_\epsilon$ in terms of the generator $\epsilon$ (see Eq. \eqref{delta H xi}). Mathematically  \cite{Wald:1993nt,Iyer:1994ys,HS:2015xlp},

{\footnotesize
\begin{equation}
\eta_{_\mathrm{H}}=\frac{1}{T_{_\mathrm{H}}}\Big(\eta_{_M}-(\Omega_{_\mathrm{H}}^i-\Omega^i_{_\infty})\eta_{_{J_i}}-(\Phi^a_{_\mathrm{H}}-\Phi^a_{_\infty})\eta_{_{Q_a}}\Big)
 \quad\Rightarrow\quad
 \delta S_{_\mathrm{H}}=\frac{1}{T_{_\mathrm{H}}}\Big(\delta M-(\Omega_{_\mathrm{H}}^i-\Omega^i_{_\infty})\delta J_i-(\Phi^a_{_\mathrm{H}}-\Phi^a_{_\infty})\delta Q_a\Big)
\end{equation}
}

\noindent where $\delta S_{_\mathrm{H}}\equiv \delta H_{\eta_{_\mathrm{H}}}\,, \,\delta M\equiv \delta H_{\eta_{_M}}\,,\,\delta J_i\equiv \delta H_{\eta_{_{J_i}}}$ and $\delta Q_a\equiv \delta H_{\eta_{_{Q_a}}}$. Notice that the $\delta$ in the proof is a generic perturbation which satisfies linearized e.o.m. So, it is not restricted to the parametric variations. Moreover, integration over horizons or asymptotics does not play any role in this proof.
\section{Applying the method to higher curvature theories}\label{sec-Higher d}
SPSM has reproduced successfully conserved charges and first law(s) for the standard (black hole) solutions to Einstein-Hilbert gravitational theories. Explicit examples can be found in Refs.  \cite{Hajian:2015eha,HS:2015xlp,Hajian:2016kxx}. The goal of this section is utilizing the SPSM for the gravitational theories with higher curvature terms. Explicitly, the Lagrangian which we will focus on, has the metric $g_{\alpha\beta}$, some gauge fields $A^a_\mu$, and some scalar fields $\phi^I$, in arbitrary dimension $d$:
\vspace*{-0.2cm}

{\small
\begin{flalign}\label{Lagrangian scalar}
\mathcal{L}=\frac{1}{16\pi G}\Big(&f(R,\phi)\!+\mathrm{a}(\phi) R_{\mu\nu}R^{\mu\nu}\!+\mathrm{b}(\phi) R_{\mu\nu\alpha\beta}R^{\mu\nu\alpha\beta}\!-\mathrm{c}_{ab}(\phi)F_{\mu\nu}^a F^{b\, \mu\nu}\!-\!2\mathrm{d}_{_{IJ}}(\phi)\nabla^\mu\phi^I\nabla_\mu\phi^J\Big).\!\!\!\!&&
\end{flalign}}

\vspace*{-0.2cm}
\noindent The $R^\mu_{\,\,\nu\alpha\beta}$, $R_{\mu\nu}$, and $R$ are Riemann tensor, Ricci tensor, and  Ricci scalar, respectively. $F^a=\mrd A^a$ are the field strengths. The coefficients $\mathrm{a}(\phi)$, $\mathrm{b}(\phi)$, $\mathrm{c}_{ab}(\phi)$, and $\mathrm{d}_{_{IJ}}(\phi)$ are some functions of $\phi^I$. Notice that the $f(R,\phi)$ covers the Einstein-Hilbert gravity with a cosmological constant. Besides, the Gauss-Bonnet theory of gravity is also covered by the Lagrangian \eqref{Lagrangian scalar}, hence the simplest Lanczos-Lovelock theories are also included  \cite{Lonczos1,Lonczos2,Lovelock,Padmanabhan:2013xyr}. Generalization to higher Lanczos-Lovelock theories is straightforward, and we will not consider in this paper. The Lagrangian $d$-form is the Hodge dual of \eqref{Lagrangian scalar}, $\mathbf{L}=\star \mathcal{L}$,
\begin{equation}\label{Lagrangian top form}
\mathbf{L}=\frac{\sqrt{-g}}{d!}\,\,\epsilon_{\mu_1\mu_2\cdots \mu_d}  \,\mathcal{L}\,\,\mrd x^{\mu_1}\wedge \mrd x^{\mu_2}\wedge\cdots\wedge \mrd x^{\mu_d}\,.
\end{equation}
The $\epsilon_{\mu_1\mu_2\cdots \mu_d}$ is the Levi-Civita symbol, i.e.  $\epsilon_{_{012\cdots d-1}}=+1$ and its sign changes with the odd permutations of indices. We will use the conventions
\begin{equation}
h^{\mu\nu}\equiv \delta g^{\mu\nu}=g^{\mu\alpha}g^{\nu\beta}\delta g_{\alpha\beta},\qquad \delta F^{\mu\nu}\equiv g^{\mu\alpha}g^{\nu\beta}(\delta \mrd A)_{\alpha\beta}=g^{\mu\alpha}g^{\nu\beta}(\mrd \delta A)_{\alpha\beta}\,.
\end{equation}
Hence, the indices for the perturbed fields can be raised and lowered similar to other tensors. Let us label the terms in the Lagrangian \eqref{Lagrangian scalar} by $f$, a, b, c, and d, respectively. The e.o.m for the chosen Lagrangian, considering variations with respect to the metric, gauge fields, and scalar fields are, respectively  \cite{Stelle:1977ry}
\vspace*{-0.2cm}

{\small \begin{flalign}
&E_{f\,\mu\nu}+E_{\text{a}\,\mu\nu}+E_{\text{b}\,\mu\nu}+E_{\text{c}\,\mu\nu}+E_{\text{d}\,\mu\nu}=0\,, \label{eom g}\\
&\hspace{2cm}E_{f\,\mu\nu}=\frac{1}{2}f g_{\mu\nu}-f' R_{\mu\nu}+\nabla_{\mu}\nabla_\nu f' -\Box f' g_{\mu\nu}\nonumber\\
&\hspace{2cm}E_{\text{a}\,\mu\nu}=\mathrm{a}\big(\frac{1}{2}R_{\alpha\beta}R^{\alpha\beta}g_{\mu\nu}+\nabla^\alpha(\nabla_\mu R_{\alpha\nu}+\nabla_\nu R_{\alpha\mu})-\nabla_\alpha\nabla_\beta R^{\alpha\beta} g_{\mu\nu}-\Box R_{\mu\nu}-2R_{\mu\alpha}R^{\alpha}_\nu\big) \nonumber\\
&\hspace{2cm}E_{\text{b}\,\mu\nu}=\mathrm{b}\big( \frac{1}{2}R_{\rho\sigma\alpha\beta}R^{\rho\sigma\alpha\beta}g_{\mu\nu}-2R_{\mu\gamma \alpha\beta}R_\nu^{\,\,\gamma \alpha\beta}-2\nabla^\alpha\nabla^\beta(R_{\mu\alpha\nu\beta}+R_{\nu\alpha\mu\beta})\big)\nonumber\\
&\hspace{2cm}E_{\text{c}\,\mu\nu}=2\text{c}_{ab}\big(F_{\mu\alpha}^a F_\nu^{b\,\alpha}-\frac{1}{4}F_{\alpha\beta}^a F^{b\, \alpha\beta}g_{\mu\nu} \big)\nonumber\\
&\hspace{2cm}E_{\text{d}\,\mu\nu}=2\text{d}_{_{IJ}}\big(\nabla_\mu \phi^I \nabla_\nu \phi^J -\frac{1}{2}\nabla^\alpha\phi^I\nabla_\alpha\phi^J g_{\mu\nu} \big)\,, \nonumber\\
&\nabla_\nu \big(\mathrm{c}_{ab} F^{b\, \mu\nu}\big)=0\,,\\
&4\nabla_\alpha \big(\mathrm{d}_{_{IJ}}\nabla^\alpha \phi^J\big)+\frac{\partial f}{\partial \phi^I} +\frac{\partial\mathrm{a}}{\partial \phi^I} R_{\mu\nu}R^{\mu\nu}\!+\frac{\partial \mathrm{b}}{\partial\phi^I} R_{\mu\nu\alpha\beta}R^{\mu\nu\alpha\beta} \!-\!\frac{\partial\mathrm{c}_{ab}}{\partial \phi^I}F_{\mu\nu}^a F^{b\, \mu\nu}\!-\!2\frac{\partial\mathrm{d}_{_{JK}}}{\partial \phi^I}\nabla^\mu\phi^J\nabla_\mu\phi^K=0, \!\!\!\!&&
\end{flalign}}
 
\vspace*{-0.2cm}
\noindent where the notation $f'\equiv \frac{\partial f}{\partial R}$ is used. We need to find  $\mathbf{\Theta}_{_\text{LW}}$, $\mathbf{Q}_\epsilon$, and most importantly, the $\boldsymbol{k}_\epsilon$ for this theory. Their derivation and final results are standard practices in the literature. Hence we only report the final results here. Detailed analysis are similar to the simple Einstein-Hilbert Lagrangian, which can be found e.g.  in Appendix A of Ref.  \cite{Hajian:2016kxx}.

By variation of Lagrangian $\delta \mathbf{L}$ and imposing the e.o.m, the surface $d\!-\!1$-form  $\mathbf{\Theta}_{_\text{LW}}$ can be read through Eq. \eqref{find Theta} to be $\mathbf{\Theta}_{_\text{LW}}=\star \Theta$, i.e. 
\begin{equation}
\mathbf{\Theta}_{_\text{LW}}=\frac{\sqrt{-g}}{(d-1)!}\,\,\epsilon_{\mu\mu_1\cdots \mu_{d-1}} \,(\Theta_{f}^\mu+\Theta_{\text{a}}^\mu+\Theta_{\text{b}}^\mu+\Theta_{\text{c}}^\mu+\Theta_{\text{d}}^\mu)\,\,\mathrm{d}x^{\mu_1}\wedge \cdots\wedge \mathrm{d}x^{\mu_{d-1}} 
\end{equation}
in which
\vspace*{-0.2cm}

{\small \begin{flalign}
&\Theta_{f}^{\mu}(\delta\Phi,\Phi)=\frac{1}{16\pi G}\big(f'(\nabla_\alpha h^{\mu\alpha}-\nabla^\mu h)-\nabla_\alpha f'h^{\mu\alpha}+\nabla^\mu f' h\big)\,, \nonumber\\
&\Theta_{\text{a}}^\mu(\delta\Phi,\Phi)=\frac{\text{a}}{16\pi G}\big(2R_{\alpha\beta}\nabla^\alpha h^{\beta \mu}-2\nabla_\alpha R^{\mu}_{\,\,\beta}h^{\alpha\beta}\!-\!R^\mu_{\,\,\alpha}\nabla^\alpha h+\nabla^\alpha R^\mu_{\,\,\alpha}h-R_{\alpha\beta}\nabla^\mu h^{\alpha\beta}\!+\!\nabla^\mu R_{\alpha\beta}h^{\alpha\beta}\big)\,,\nonumber\\
&\Theta_{\text{b}}^\mu(\delta\Phi,\Phi)=\frac{\text{b}}{4\pi G}( \nabla^\nu R^{\mu}_{\,\,\alpha\nu\beta}h^{\alpha\beta}-R^{\mu}_{\,\,\alpha\nu\beta}\nabla^\nu h^{\alpha\beta})\,,\nonumber\\
&\Theta_{\text{c}}^\mu(\delta\Phi,\Phi)=\frac{-1}{4\pi G}\mathrm{c}_{ab}\,F^{a\,\mu\nu}\,\delta A^b_{\nu}\,,\nonumber\\
&\Theta_{\text{d}}^\mu(\delta\Phi,\Phi)=\frac{-1}{4\pi G}\mathrm{d}_{_{IJ}}\,\nabla^\mu\phi^I\delta\phi^J\,,&&
\end{flalign}}

\vspace*{-0.2cm}
\noindent where $h\equiv h^\alpha _{\,\,\alpha}$.  Having the $\mathbf{\Theta}$ in our hand, for a generic $\epsilon=\{\xi,\lambda^a\}$, the Noether-Wald $d\!-\!2$-form $\mathbf{Q}_\epsilon$ can be read through Eq. \eqref{Noether Wald} and imposing the e.o.m Eq. \eqref{eom g}, as 
\begin{equation}
\mathbf{Q}_\epsilon=\frac{\sqrt{-g}}{(d-2)!\,2!}\,\,\epsilon_{\mu\nu\mu_1\cdots \mu_{d-2}} \,(\mathrm{Q}_{f\,\epsilon}^{\mu\nu}+\mathrm{Q}_{\mathrm{a}\,\epsilon}^{\mu\nu}+\mathrm{Q}_{\mathrm{b}\,\epsilon}^{\mu\nu}+\mathrm{Q}_{\text{c}\,\epsilon}^{\mu\nu}+\mathrm{Q}_{\mathrm{d}\,\epsilon}^{\mu\nu})\,\,\mathrm{d}x^{\mu_1}\wedge \cdots\wedge \mathrm{d}x^{\mu_{d-2}} 
\end{equation}
 in which 
\vspace*{-0.3cm}

{\small
\begin{flalign}
&\mathrm{Q}_{f\,\epsilon}^{\mu\nu}=\frac{1}{16\pi G}\big(2\nabla^\mu f' \xi^\nu-f' \nabla^\mu \xi^\nu \big)- [\mu\leftrightarrow\nu]\,,\nonumber\\
&\mathrm{Q}_{\text{a}\,\epsilon}^{\mu\nu}=\frac{\text{a}}{8\pi G}\big(\nabla^\mu R^\nu_{\,\,\alpha}\xi^\alpha+ R^{\nu}_{\,\,\alpha}\nabla^\alpha\xi^\mu - \nabla^\alpha R^\nu_{\,\,\alpha}\xi^\mu\big)-[\mu\leftrightarrow\nu]\,,\nonumber\\
&\mathrm{Q}_{\text{b}\,\epsilon}^{\mu\nu}=\frac{\text{b}}{4\pi G}\big(\nabla^\alpha R^{\mu\nu}_{\,\,\,\,\,\,\alpha\beta}\,\xi^\beta-R^{\mu\alpha\nu\beta}\nabla_\alpha\xi_\beta\big)-[\mu\leftrightarrow\nu]\,,\nonumber\\
& \mathrm{Q}_{\text{c}\,\epsilon}^{\mu\nu}=\frac{-1}{4\pi G}\mathrm{c}_{ab}F^{a\,\,\mu\nu}(A^b_\rho\xi^\rho+\lambda^b)\,,\nonumber\\
&\mathrm{Q}_{\text{d}\,\epsilon}^{\mu\nu}=0\,.&&
\end{flalign}}

\vspace*{-0.2cm}
\noindent After varying the $\mathbf{Q}_\epsilon$ with respect to all dynamical fields and utilizing the textbook relations
\vspace*{-0.3cm}

{\small \begin{flalign}
&\delta \sqrt{-g}=\frac{\sqrt{-g}}{2}h^\alpha_{\,\,\alpha}\,, \quad \quad \qquad \delta \Gamma ^\lambda_{\mu\nu}= \frac{1}{2}g^{\lambda \sigma}(\nabla _\mu h_{\sigma \nu}+\nabla_\nu h_{\sigma \mu}-\nabla _\sigma h_{\mu\nu}),\quad \qquad \qquad \delta \epsilon_{\mu\nu\mu_1\cdots \mu_{d-2}}=0\,,\nonumber\\
&\delta R_{\mu\nu\alpha\beta}=\frac{1}{2}\big(2R_{\mu\nu\alpha\gamma}\,h^\gamma_{\,\,\beta}-\nabla_\mu\nabla_\alpha h_{\beta\nu}\!+\!\nabla_\mu\nabla_\beta h_{\alpha\nu}\!-\!\nabla_\mu\nabla_\nu h_{\alpha\beta}\!+\!\nabla_\nu\nabla_\alpha h_{\beta\mu}\!-\!\nabla_\nu\nabla_\beta h_{\alpha\mu}\!+\!\nabla_\nu\nabla_\mu h_{\alpha\beta}\big),\nonumber\\
&\delta R_{\mu\nu}=\frac{1}{2}\left( \nabla_\alpha \nabla_\mu h^\alpha_{\,\,\nu}+\nabla_\alpha \nabla_\nu h^\alpha_{\,\,\mu}-\Box h_{\mu\nu}-\nabla_\mu\nabla_\nu h \right),   \qquad \delta R=\nabla_\mu \nabla_\nu h^{\mu\nu}-\Box h-R_{\mu\nu}h^{\mu\nu}\,,&&
\end{flalign}}

\vspace*{-0.2cm}
\noindent one can find  $\boldsymbol{k}_\epsilon$ by \eqref{k_xi general}, to calculate variations of the conserved charges. The result, the final applicable tensor for calculation of conserved charges turns out to be
\begin{equation}\label{k epsilon general}
\boldsymbol{k}_\epsilon(\delta\Phi,\Phi)=\frac{\sqrt{-g}}{(d-2)!\,2!}\,\,\epsilon_{\mu\nu\mu_1\cdots \mu_{d-2}} \,(k_{f\,\epsilon}^{\mu\nu}+k_{\text{a}\,\epsilon}^{\mu\nu}+k_{\text{b}\,\epsilon}^{\mu\nu}+k_{\text{c}\,\epsilon}^{\mu\nu}+k_{\text{d}\,\epsilon}^{\mu\nu})\,\,\mathrm{d}x^{\mu_1}\wedge \cdots\wedge \mathrm{d}x^{\mu_{d-2}} 
\end{equation}
where, using the notations ${\small f'\equiv \frac{\partial f}{\partial R}}$,
\vspace*{-0.2cm}

{\small
\begin{flalign}
&k_{f\,\epsilon}^{\mu\nu}(\delta\Phi,\Phi)=\dfrac{1}{16 \pi G}\Big[\Big( h^{\mu\alpha}\nabla_\alpha\xi^\nu-\nabla^\mu h^{\nu\alpha}\xi_\alpha-\frac{1}{2}h \nabla^\mu\xi^\nu\Big)f'+2\Big(R^{\mu\alpha}\nabla_\alpha h-\nabla_\alpha R h^{\mu\alpha}-R^\mu_{\,\,\alpha}\nabla_\beta h^{\alpha\beta}\nonumber\\
&\hspace*{2.9cm}-\Box \nabla^\mu h +\nabla_\alpha\nabla^\mu\nabla_\beta h^{\alpha\beta}-\nabla^\mu(R_{\alpha\beta} h^{\alpha\beta})+\frac{1}{2}\nabla^\mu R\, h \Big)\xi^\nu f''\!\!\nonumber\\
&\hspace*{2.9cm}+\!2(\nabla^\mu \delta \phi^I \!-\! h^\mu_{\,\,\alpha}\nabla^\alpha\phi^I+\frac{1}{2}h\nabla^\mu\phi^I)\xi^\nu\!\frac{\partial f'}{\partial\phi^I}- \delta \phi^I \nabla^\mu\xi^\nu \frac{\partial f'}{\partial \phi^I} \nonumber\\
&\hspace*{2.9cm}+\Big(R_{\alpha\beta}h^{\alpha\beta}-\nabla_\alpha\nabla_\beta h^{\alpha\beta}
+\Box h\Big)(\nabla^\mu\xi^\nu f''-2\nabla^\mu R \,\xi^\nu f'''-2\nabla^\mu\phi^I\, \xi^\nu \frac{\partial f''}{\partial \phi^I})\nonumber\\
&\hspace*{2.9cm}+2\delta \phi^I \nabla^\mu \phi^J\, \xi^\nu \frac{\partial^2 f'}{\partial\phi^I\partial \phi^J}+2\delta \phi^I \nabla^\mu R \,\xi^\nu \frac{\partial f''}{\partial \phi^I} \Big]-\Theta^\mu_f \xi^\nu-[\mu\leftrightarrow\nu], \\
\hspace*{0.2cm}\nonumber\\
&k_{\text{a}\,\epsilon}^{\mu\nu}(\delta\Phi,\Phi)=\dfrac{\text{a}}{16 \pi G}\Big[\Big(\nabla^\alpha R_\alpha^{\,\,\mu} h-\nabla_\alpha R h^{\mu\alpha}\!-\!\nabla^\mu(R_{\alpha\beta}h^{\alpha\beta})+\nabla^\mu\nabla_\alpha\nabla_\beta h^{\alpha\beta}-\nabla^\mu \Box h\Big)\xi^\nu \!+\! \Big( 2\nabla_\beta R^\mu_{\,\,\alpha}h^{\beta\nu}\nonumber\\
&\hspace*{2.9cm}-2 R^{\mu\beta}\nabla_\beta h^\nu_{\,\, \alpha}-2\nabla^\mu R_{\alpha\beta}h^{\nu\beta}- \nabla^\mu (\nabla_\alpha\nabla^\nu h -\nabla_\beta \nabla_\alpha h^{\nu\beta} +\Box h^\nu_{\,\,\alpha}-\nabla^\beta \nabla^\nu h_{\alpha\beta}) \nonumber\\
&\hspace*{2.9cm} +\nabla^\mu R^\nu_{\,\,\alpha} h +2 R^{\mu\beta} \nabla^\nu h_{\alpha\beta}\Big) \xi^\alpha +\Big(\nabla_\alpha \nabla^\mu h - \nabla_\beta \nabla_\alpha h^{\mu\beta}-\nabla^\beta \nabla^\mu h_{\alpha\beta}+\Box h^{\mu}_{\,\,\alpha}\nonumber\\
&\hspace*{2.9cm} +2(R_{\alpha\beta}h^{\mu\beta}+R^{\mu\beta}h_{\alpha\beta})-R^{\mu}_{\,\,\alpha} h\Big) \nabla^\alpha \xi^\nu \Big]\!+\!\frac{\mathrm{Q}_{\text{a}\,\epsilon}^{\mu\nu}}{2\text{a}}\frac{\partial \text{a}}{\partial\phi^I}\delta \phi^I\!-\!\Theta^\mu_{\text{a}} \xi^\nu\!-\![\mu\leftrightarrow\nu],\!\!\!\!\! 
\end{flalign} 
\begin{flalign}
&k_{\text{b}\,\epsilon}^{\mu\nu}(\delta\Phi,\Phi)=\dfrac{\text{b}}{8 \pi G}\Big[\Big(2(R^\mu_{\,\,\alpha\beta\gamma}\!\!-\!R^\mu_{\,\,\beta\alpha\gamma})h^{\nu\gamma}\!\!+\! R^{\mu\,\,\,\nu}_{\,\, \alpha\,\, \beta}h \!-\! R^{\mu\,\,\,\nu}_{\,\,\alpha\,\,\gamma}h_\beta^{\,\,\gamma}\!-\! R^{\mu\,\,\,\nu}_{\,\,\beta\,\,\gamma}h_\alpha^{\,\,\gamma} \!-\! \nabla^\mu \nabla_\alpha h^\nu_{\,\,\beta}\!+\!\nabla^\mu \nabla_\beta h^\nu_{\,\,\alpha}\Big)\!\nabla^\beta\xi^\alpha\nonumber\\
&\hspace*{2.9cm} +\Big(R^{\mu\beta}(\nabla_\beta h^\nu_{\,\, \alpha}-\nabla_\alpha h^\nu_{\,\,\beta})+R^{\mu\,\,\,\nu}_{\,\,\beta\,\,\gamma} \nabla^\gamma h_\alpha^{\,\,\beta}+\frac{1}{2}R^{\mu\nu}_{\,\,\,\,\,\alpha\gamma}(\nabla_\beta h^{\beta\gamma}-\nabla^\gamma h)\nonumber\\
&\hspace*{2.9cm} +2(\nabla_\beta R^\mu_{\,\,\alpha}-\nabla^\mu R_{\alpha\beta})h^{\nu\beta}+\nabla^\mu \nabla_\beta\nabla_\alpha h^{\nu\beta}-\nabla^\mu\Box h^\nu_{\,\,\alpha}+\nabla^\mu R^\nu_{\,\,\alpha}h + \nabla^\mu R^\nu_{\,\,\beta} h_\alpha^{\,\,\beta}\nonumber\\
&\hspace*{2.9cm} -\nabla^\mu (R^\nu _{\,\,\beta\alpha\gamma}h^{\beta\gamma})\Big) 2\xi^\alpha\Big]+\frac{\mathrm{Q}_{\text{b}\,\epsilon}^{\mu\nu}}{2\text{b}}\frac{\partial \text{b}}{\partial\phi^I}\delta \phi^I -\Theta^\mu_{\text{b}} \xi^\nu-[\mu\leftrightarrow\nu], \\
&k_{\text{c}\,\epsilon}^{\mu\nu}(\delta\Phi,\Phi)=\frac{1}{8 \pi G}\Big[\big(\frac{-h}{2} \,\mathrm{c}_{ab}\, F^{a\,\mu\nu}\!+\!2\,\mathrm{c}_{ab}\,F^{a\,\mu\sigma}h_\sigma^{\;\;\nu}-\mathrm{c}_{ab}\,\delta F^{a\,\mu\nu}\!-\!\frac{\partial\,\mathrm{c}_{ab}}{\partial \phi^I}\,F^{a\,\mu\nu}\delta\phi^I\big)({\xi}^\alpha A^b_\alpha+\lambda^b)-\nonumber\\
 & \hspace*{2.9cm}\,\mathrm{c}_{ab}\,F^{a\,\mu\nu}\xi^\alpha \delta A^b_\alpha-2\,\mathrm{c}_{ab}\,F^{a\,\alpha\mu}\xi^\nu \delta A^b_\alpha\Big]-[\mu\leftrightarrow\nu]\,, \\
 &k_{\text{d}\,\epsilon}^{\mu\nu}(\delta\Phi,\Phi)=\frac{1}{4\pi G}\Big[\xi^\nu\,\mathrm{d}_{_{IJ}}\,\nabla^\mu\phi^I\,\delta\phi^J\Big]-[\mu\leftrightarrow\nu]\,. &&
\end{flalign} 
}

\vspace*{-0.2cm}
\noindent Having $\boldsymbol{k}_\epsilon$, and equipped with the parametric variations $\hat \delta \Phi$, calculation of the conserved charges associated with the exact symmetries $\eta=\{\zeta,\lambda^a\}$ of  the (black hole) solutions $\hat \Phi(x^\mu;p_j)$ to the Lagrangian \eqref{Lagrangian scalar} can be performed.  

\section{Some examples}\label{sec-example}
To exemplify, in this section we will work out conserved charges and first law(s) of thermodynamics for some black hole solutions to the Lagrangian \eqref{Lagrangian scalar}. \\

\noindent\textbf{Example 1: $\mathbf{z=3}$ Lifshitz black hole in $\boldsymbol{d}\mathbf{=3}$}\\
\noindent Consider
\begin{equation}\label{coefficients ex 1}
f=R+\frac{13}{l^2}-\frac{3l^2}{4}R^2, \qquad \text{a}=2l^2, \qquad \text{b}=\text{c}=\text{d}=0\,,
\end{equation}
i.e.  the new massive gravity (NMG) Lagrangian \cite{Bergshoeff:2009hq}
\begin{equation}\label{NMG}
\mathcal{L}=\frac{1}{16\pi G}\left(R-2\Lambda+\frac{1}{\mathfrak{m}^2}(R_{\mu\nu}R^{\mu\nu}-\frac{3}{8}R^2)\right),
\end{equation}
in which $\Lambda=-\frac{13}{2l^2}$ and $\mathfrak{m}^2=\frac{1}{2l^2}$. We can have a family of black holes $\hat g_{\alpha\beta}(x^\mu;m)$ as solution to this theory in 3-dimensional case \cite{AyonBeato:2009nh,AyonBeato:2010tm}
\begin{equation}
\mrd s^2=-(\frac{r}{l})^{2z}(1-\frac{ml^2}{r^2})\,\mrd t^2 +\frac{\mrd r^2}{\frac{r^2}{l^2}(1-\frac{ml^2}{r^2})}+r^2 \mrd \varphi^2
 \end{equation} 
for the $z=3$. Let us analyze the thermodynamics of this family of black holes using SPSM. Putting Eq. \eqref{coefficients ex 1} into the general result  Eq. \eqref{k epsilon general}, $k_\epsilon^{\mu\nu}$ can be read for our specific theory.  Then, choosing $\partial \Sigma$ to be surfaces of constant $(t,r)$ for simplicity, the conserved charge variations for an exact symmetry $\eta$ can be simply read through
\begin{align}\label{k epsilon tr}
\hat\delta H_{\eta}= \oint_{\partial \Sigma}\boldsymbol{k}_\eta(\hat\delta g_{\alpha\beta},\hat g_{\alpha\beta}) =\int_0^{2\pi}\sqrt{-\hat g}\,k_\eta^{tr}(\hat\delta g_{\alpha\beta},\hat g_{\alpha\beta}) \,\mathrm{d}\varphi\,,
\end{align}
in which $k_\eta^{tr}$ is the $tr$ component of the $k_\eta^{\mu\nu}$. Inserting parametric variations $\hat\delta g_{\alpha\beta}=\frac{\partial \hat g_{\alpha\beta}}{\partial m}\delta m$ in it, conserved charges can be calculated, irrespective of the asymptotic Lifshitz behavior.\\
\underline{Mass}: We can choose the stationarity Killing $-\partial_t$ as the generator to which the mass is associated. The minus sign has been adopted to make the mass and entropy positive. Hence, by $\eta_{_M}=\{-\partial_t,0\}$ the result of calculating Eq. \eqref{k epsilon tr} is
\begin{equation}\label{hat del M 1}
\hat \delta M\equiv \hat \delta H_{\eta_{_M}}=\frac{m}{2G}\delta m= \hat \delta(\frac{m^2}{4G})  \qquad \Rightarrow \qquad M=\frac{m^2}{4G}.
\end{equation}
The reference point (constant of integration) was chosen $M\!=\!0$ for the geometry with $m=0$.\\
\noindent\underline{Angular momentum}: Choosing $\eta_{_J}=\{-\partial_\varphi,0\}$, and by a similar analysis as the mass, angular momentum turns out to be
\begin{equation}
\hspace*{0.84cm}\hat \delta J\equiv \hat \delta H_{\eta_{_J}}=0\times \delta m \qquad \Rightarrow \qquad J=0.
\end{equation}
\noindent\underline{Entropy}: The surface gravity on the horizon of this solution is $\kappa_{_\mathrm{H}}=\frac{r_{_\mathrm{H}}^3}{l^4}$ in which $r_{_\mathrm{H}}=\sqrt{ml^2}$. The entropy of the horizon is defined to be the conserved charge associated with the horizon Killing vector $\zeta_{_\mathrm{H}}$ normalized by the Hawking temperature $T_{_\mathrm{H}}=\frac{\kappa_{_\mathrm{H}}}{2\pi}$. Therefore, by $\eta_{_\mathrm{H}}=\frac{2\pi}{\kappa_{_\mathrm{H}}}\{\zeta_{_\mathrm{H}},0\}$ and the identity $\zeta_{_\mathrm{H}}= -\partial_t$, the entropy attributed to the horizon, via a similar integration to the other conserved charges, is calculated to be
\begin{equation}\label{hat del ent 1}
\hspace*{-0.54cm}\hat \delta S_{_\mathrm{H}}\equiv \hat \delta H_{\eta_{_\mathrm{H}}}=\frac{\pi l}{G \sqrt{m}}\delta m=\hat \delta (\frac{2\pi r_{_\mathrm{H}}}{G}) \qquad \Rightarrow\qquad     S_{_\mathrm{H}}=\frac{2\pi r_{_\mathrm{H}}}{G}\,.
\end{equation}
The reference point is chosen to be $S_{_\text{H}}\!=\!0$ for the geometry identified by $m=0$.
Notice that the entropy is proportional to the area (here the length) of the horizon, but without the usual factor of $\frac{1}{4}$. The results above are in agreement with the results reported in Refs.  \cite{Hohm:2010jc,Gim:2014nba}.\\
\noindent\underline{First law}: Having made the entropy free of being calculated on the horizons, the first law of thermodynamics would follow:
\begin{equation}
\eta_{_\mathrm{H}}=\frac{1}{T_{_\mathrm{H}}}\eta_{_M}
\quad \xrightarrow{\text{linearity of $\delta H_\epsilon$ in $\epsilon$\,\,\,}}
\qquad \delta S_{_\mathrm{H}}=\frac{1}{T_{_\mathrm{H}}}\delta M\,.
\end{equation}
Although $\delta$ in the equation above is a generic perturbation which satisfies linearized e.o.m, but it can be cross-checked for the parametric variations using the explicit results for $\hat \delta M$ and $\hat \delta S_{_\mathrm{H}}$ in Eqs. \eqref{hat del M 1} and \eqref{hat del ent 1}.\\

\noindent\textbf{Example 2: Warped AdS$_3$}\\
A warped AdS$_3$ is a $3$-dimensional black hole identified by two parameters $p_{_1}=m$ and $p_{_2}=j$: 

{
\begin{align}\label{WBTZ}
&\mrd s^2= \Big( \frac{-r^2}{l^2}+8(m-\frac{j}{l})\Big) \mrd t^2+\frac{\mrd r^2}{\frac{16 j^2}{r^2}+\frac{r^2}{l^2}-8(m-\frac{j}{l})}+r^2\mrd \varphi^2 -(\omega_t \mrd t-\omega_\varphi \mrd \varphi)^2\nonumber\\
& \omega_t\equiv\frac{H(-r^2+8l^2m-4lj)}{2l^2\sqrt{m}}\,,\qquad \qquad \omega_\varphi\equiv\frac{H(r^2+4lj)}{2l\sqrt{m}}\,.
\end{align}
}

\noindent It is a solution \cite{Clement:2009gq,Detournay:2015ysa} to the NMG theory, described by the Lagrangian in Eq. \eqref{NMG} with
\begin{equation}
\Lambda=\frac{84\,H^4+60\,H^2-35}{2\,l^2(17-42\,H^2)}, \qquad \frac{1}{\mathfrak{m}^2}=\frac{2\,l^2}{42\,H^2-17}\,.
\end{equation}
After extracting $k^{\mu\nu}_\epsilon$ for this theory from the general result in Eq. \eqref{k epsilon general}, and equipped with the parametric variations
\begin{equation}
\hat\delta g_{\alpha\beta}=\frac{\partial \hat g_{\alpha\beta}}{\partial m}\delta m+\frac{\partial \hat g_{\alpha\beta}}{\partial j}\delta j\,,
\end{equation}
 one can find the conserved charges by an integration similar to the Eq. \eqref{k epsilon tr}. Notice that because of the linearity of $\delta H_\eta (\delta \Phi,\Phi)$ in $\delta \Phi$, the parametric variations can be inserted term by term into the calculations. This makes the calculations to be performed easier. \\ 
\underline{Mass}: By $\eta_{_M}=\{\partial_t+\Omega_{_\infty}\partial_\varphi,0\}$ in which $\Omega_{_\infty}=\frac{-1}{l}$, it turns out that
\begin{equation}\label{hat del M 2}
\hat \delta M\equiv \hat \delta H_{\eta_{_M}}=\frac{16 (1-2H^2)^{\frac{3}{2}}}{G(17-42\,H^2)}\delta m + 0\times \delta j  \qquad \Rightarrow \qquad M=\frac{16 (1-2H^2)^{\frac{3}{2}}m}{G(17-42\,H^2)}.
\end{equation}
\noindent\underline{Angular momentum}: Choosing $\eta_{_J}=\{-\partial_\varphi,0\}$, 
\begin{equation}\label{hat del J 2}
\hspace*{-0.4cm}\hat \delta J\equiv \hat \delta H_{\eta_{_J}}=0\times \delta m+ \frac{16 (1-2H^2)^{\frac{3}{2}}}{G(17-42\,H^2)}\delta j\qquad \Rightarrow \qquad J=\frac{16 (1-2H^2)^{\frac{3}{2}}j}{G(17-42\,H^2)}.
\end{equation}
\noindent\underline{Entropies}: There are two horizons in the warped AdS$_3$ geometry \eqref{WBTZ}. So, we would find two entropies attributed to them. The horizons are situated at $r_\pm^2=4l^2(m-\frac{j}{l}\pm \sqrt{m(m-\frac{2j}{l}})$, collectively denoted by $r_{_\mathrm{H}}$. The surface gravities, angular velocities, and the Killing vectors of the horizons are
\begin{equation}
\kappa_{_\mathrm{H}}=\frac{r_{_\mathrm{H}}^4-16\, l^2j^2}{l^2 r_{_\mathrm{H}}^3}\,, \qquad \Omega_{_\mathrm{H}}=\frac{4j}{r_{_\mathrm{H}}^2}\,, \qquad \zeta_{_\mathrm{H}}=\partial_t+\Omega_{_\mathrm{H}}\partial_\varphi\,,
\end{equation}
respectively. Integrating over arbitrary surfaces of constant time and radius, the entropies as conserved charges associated with the exact symmetries $\eta_{_\mathrm{H}}=\frac{2\pi}{\kappa_{_\mathrm{H}}}\{\zeta_{_\mathrm{H}},0\}$ are calculated to be 
\begin{equation}\label{hat del S 2}
\hat \delta S_{_\mathrm{H}}= \frac{\partial (\frac{8\pi (1-2H^2)^{\frac{3}{2}}r_{_\mathrm{H}}}{G(17-42\,H^2)})}{\partial m}\delta m+\frac{\partial (\frac{8\pi (1-2H^2)^{\frac{3}{2}}r_{_\mathrm{H}}}{G(17-42\,H^2)})}{\partial j}\delta j \qquad \Rightarrow\qquad     S_{_\mathrm{H}}=\frac{8\pi (1-2H^2)^{\frac{3}{2}}r_{_\mathrm{H}}}{G(17-42\,H^2)}\,.
\end{equation}
The reference point for all of the charges above have been chosen to vanish on the geometry identified by $m=j=0$. Our results would match exactly with the results reported in Refs.  \cite{Ghodrati:2016ggy,Detournay:2016gao} if one replaces the parameter $m\to m-\frac{j}{l}$. The difference originates from considering the asymptotic angular velocity $\Omega_{_\infty}$in the definition of mass. Hence, the mass calculated here is different from the mass reported in  \cite{Ghodrati:2016ggy,Detournay:2016gao} by a term, which is $\Omega_{_\infty}J$. \\
\noindent\underline{First laws}: For any generic perturbation which satisfies the linearized e.o.m, the first laws follow:
\begin{equation}
\eta_{_\mathrm{H}}=\frac{1}{T_{_\mathrm{H}}}(\eta_{_M}-(\Omega_{_\mathrm{H}}-\Omega_{_\infty})\eta_{_J})
\quad \xrightarrow{\text{linearity of $\delta H_\epsilon$ in $\epsilon$\,\,\,}}
\quad \delta S_{_\mathrm{H}}=\frac{1}{T_{_\mathrm{H}}}(\delta M-(\Omega_{_\mathrm{H}}-\Omega_{_\infty})\delta J)\,,
\end{equation}
which can be checked for the parametric variations in \eqref{hat del M 2}, \eqref{hat del J 2}, and \eqref{hat del S 2}.\\

\noindent\textbf{Example 3: Schwarzschild-AdS black holes in $\boldsymbol{d}$-dimensions}\\
The family of black holes
\begin{equation}\label{example 2 metric}
\mrd s^2=-(1-\frac{2Gm}{r^{d-3}}+\frac{r^2}{l^2})\mrd t^2+\frac{\mrd r^2}{1-\frac{2Gm}{r^{d-3}}+\frac{r^2}{l^2}}+r^2 \mrd \Omega_{_{d-2}}^2
\end{equation}
are solutions to the theories
\begin{equation}\label{ex 2 Lagrangian}
\mathcal{L}=\frac{1}{16\pi G}\Big(R-2\Lambda+\alpha R^2+\text{a} R_{\mu\nu}R^{\mu\nu}\Big),
\end{equation}
where $\alpha$ and $\text{a}$ are arbitrary constants, and $\Lambda =\frac{-l^2(d^2-3d+2)+(\alpha d + \text{a})(d-4)(d-1)^2}{2l^4}$. For these theories, $k^{\mu\nu}_\epsilon$ can be read through the general result \eqref{k epsilon general}
by putting $f=R-2\Lambda+\alpha R^2$, the arbitrary constant factor $\text{a}$, and vanishing  $\text{b}=\text{c}_{ab}=\text{d}_{IJ}=0$. Similar to the previous examples, one can choose $\partial \Sigma$ to be surfaces of constant $(t,r)$ for simplicity. Hence, the conserved charge variations for an exact symmetry $\eta$ would be
\begin{align}\label{k epsilon tr example 2}
\hat\delta H_{\eta}= \int\displaylimits_{S_{d-2}}\!\!\sqrt{-\hat g}\,k_\eta^{tr}(\hat\delta g_{\alpha\beta},\hat g_{\alpha\beta}).
\end{align}
The integration is taken over the $d-2$ dimensional spheres, e.g.  in four dimensions it is $\int_0^\pi\int_0^{2\pi} \mrd \theta\mrd \varphi$. By parametric variations $\hat\delta g_{\alpha\beta}=\frac{\partial \hat g_{\alpha\beta}}{\partial m}\delta m$, conserved charges can be calculated.\\
\underline{Mass}: For the exact symmetry $\eta_{_M}=\{\partial_t,0\}$, the result of calculating Eq. \eqref{k epsilon tr example 2} is
\begin{equation}\label{ex 2 M}
\hat \delta M = \mathcal{X}\times \frac{(d-2)\Omega_{_{d-2}}}{8\pi}\delta m  \qquad \Rightarrow \qquad M=\mathcal{X}\times \frac{(d-2)\Omega_{_{d-2}}}{8\pi} m\,,
\end{equation}
where 
\begin{equation}
\mathcal{X}=\frac{l^2- 2d(d-1)\alpha- 2(d-1) \text{a}}{l^2}\,, \qquad \Omega_{_{d-2}}=\frac{2\pi^{\frac{d-1}{2}}}{\Gamma(\frac{d-1}{2})}\,.
\end{equation}
The reference point has been chosen to be $M\!=\!0$ for the geometry which is identified by $m=0$.\\
\noindent\underline{Angular momentum}: By $\eta_{_J}=\{-\partial_\varphi,0\}$, angular momentum is calculated to be
\begin{equation}
\hspace*{-1cm}\hat \delta J=0\times \delta m \qquad \Rightarrow \qquad J=0.
\end{equation}
\noindent\underline{Entropy}: Surface gravity is a property of solutions, and it is independent of the theory. For the event horizon of the solutions \eqref{example 2 metric}, it is 
\begin{equation}
\kappa_{_\mathrm{H}}=\frac{(d-1)r_{_\text{H}}^{d-2}+(d-3)l^2\,r_{_\text{H}}^{d-4}}{2l^2\,r_{_\text{H}}^{d-3}},
\end{equation}
in which $r_{_\mathrm{H}}$ solves the equation $r_{_\text{H}}^{d-1}+l^2 r_{_\text{H}}^{d-3}-2Gml^2=0$. By $\eta_{_\mathrm{H}}=\frac{2\pi}{\kappa_{_\mathrm{H}}}\{\zeta_{_\mathrm{H}},0\}$ in which $\zeta_{_\mathrm{H}}= \partial_t$, the entropy variation attributed to the event horizon, which is  calculated on arbitrary surfaces of integration, would be
\begin{equation}\label{hat del S 3}
\hat\delta S_{_\text{H}}= \mathcal{X}\times \frac{(d-2)\Omega_{_{d-2}}}{4\kappa_{_\text{H}}}\delta m\,.
\end{equation}
Noticing the linearity of $\delta H_\epsilon$ \eqref{delta H xi} in $\epsilon$, this result can also be found by multiplication of $\hat \delta M$, which is calculated in Eq. \eqref{ex 2 M}, by the factor $\frac{2\pi}{\kappa_{_\text{H}}}$. Hence, using $\frac{\partial r_{_\text{H}}}{\partial m}=\frac{2Gl^2}{(d-1) r_{_\text{H}}^{d-2}+(d-3)l^2 r_{_\text{H}}^{d-4}}$
\begin{equation}
\hat\delta S_{_\text{H}}=\frac{\partial (\frac{\mathcal{X}\Omega_{_{d-2}}\,r_{_\text{H}}^{d-2}}{4G})}{\partial r_{_\text{H}}}\,\frac{\partial r_{_\text{H}}}{\partial m} \,\delta m =\hat \delta (\mathcal{X}\times \frac{\Omega_{_{d-2}}\,r_{_\text{H}}^{d-2}}{4G})\qquad \Rightarrow\qquad S_{_\text{H}}=\mathcal{X}\times \frac{\Omega_{_{d-2}}\,r_{_\text{H}}^{d-2}}{4G}\,.
 \end{equation} 
\noindent\underline{First law}: It is simply
\begin{equation}
\eta_{_\mathrm{H}}=\frac{1}{T_{_\mathrm{H}}}\eta_{_M}\\
\quad \xrightarrow{\text{linearity of $\delta H_\epsilon$ in $\epsilon$\,\,\,}}
\qquad \delta S_{_\mathrm{H}}=\frac{1}{T_{_\mathrm{H}}}\delta M\,.
\end{equation}\\
which can be checked for the parametric variations by the results in Eqs.  \eqref{ex 2 M} and \eqref{hat del S 3}.
\noindent We finish this example by mentioning two remarks:
\begin{itemize}
\item [--] One can add any factor of Gauss-Bonnet Lagrangian $\mathcal{L}_{_\text{GB}}\propto R^2-4R_{\mu\nu}R^{\mu\nu}+R_{\mu\nu\alpha\beta}R^{\mu\nu\alpha\beta}$ in $d=3,4$ to the Lagrangian \eqref{ex 2 Lagrangian}, without affecting the e.o.m \emph{and} conserved charges. 
\item [--]  The family of black holes in this example has the property $R_{\mu\nu}=\frac{R}{d}g_{\mu\nu}$. The geometries with such a property are called \emph{Einstein geometries}. The following theorem (see Ref. \cite{Mozaffar:2016hmg} and references therein) sheds light on the results of the calculations above.\\
\textit{Theorem:} Any theory which is described by a Lagrangian $\mathcal{L}=\mathcal{L}(g_{\alpha\beta},R_{\mu\nu})$, with an Einstein geometry $g_{\alpha\beta}$ as a solution, can be mapped to the Einstein-Hilbert theory with the solution
\begin{equation}
 \bar g_{\alpha\beta}=\mathcal{X}^{\frac{2}{d-2}} g_{\alpha\beta}\,, \qquad  \mathcal{X}=\Big[\frac{d}{2R}\times \mathcal{L}\Big]_{\text{on-shell}}\,.
\end{equation}  
This theorem clarifies the observation that the mass and entropy calculated above are the mass and entropy in the Einstein-Hilbert theory multiplied by the factor $\mathcal{X}$. 
\end{itemize}
\noindent\textbf{Example 4: Charged static BTZ black hole}\\
Our last example, although it is in the context of the Lagrangian \eqref{Lagrangian scalar}, but does not have higher curvature terms. It would be a pedagogical example in the presence of the gauge fields. Moreover, it remedies the divergent results appearing in the literature. This last example is the electrically charged static BTZ black hole  \cite{Banados:1992wn,Martinez:1999qi}
\begin{align}\label{ex 3 fields}
&\mrd s^2=-(-Gm+\frac{r^2}{l^2}-\frac{q^2}{2}\log{\frac{r}{l}})\mrd t^2+\frac{\mrd r^2}{-Gm+\frac{r^2}{l^2}-\frac{q^2}{2}\log{\frac{r}{l}}}+r^2 \mrd \varphi^2\nonumber\\
&\hat A=-\frac{q}{2}\log (\frac{r}{l})\, \mrd t
\end{align}
as a solution to the theory described by 
\begin{equation}
\mathcal{L}=\frac{1}{16\pi G}(R-2\Lambda-F_{\mu\nu}F^{\mu\nu})
\end{equation}
for $\Lambda=\frac{-1}{l^2}$. $k^{\mu\nu}_\epsilon$ for this theory can be read through  Eq. \eqref{k epsilon general}
by putting $f=R-2\Lambda$, $\text{a}=\text{b}=\text{d}_{IJ}=0$, and $\text{c}_{ab}=\delta_{ab}$. Making the simplifying choice of taking $\partial \Sigma$ to be the lines of constant $(t,r)$, conserved charge variations for an exact symmetry $\eta$ would be
\begin{align}\label{k epsilon tr example 3}
\hat\delta H_{\eta}= \int_0^{2\pi} \sqrt{-\hat g}\,k_\eta^{tr}(\hat\delta \Phi,\hat \Phi)\, \mrd \varphi.
\end{align} 
The dynamical fields $\hat\Phi$ are the metric $\hat g_{\alpha\beta}$ and gauge field $\hat A$ in Eq. \eqref{ex 3 fields}, parametrized by $p_j=\{m,q\}$. So, the parametric variations would be
\begin{equation}
\hat\delta g_{\alpha\beta}=\frac{\partial \hat g_{\alpha\beta}}{\partial m}\delta m+\frac{\partial \hat g_{\alpha\beta}}{\partial q}\delta q\,, \qquad \hat\delta A_{\mu}=\frac{\partial \hat A_{\mu}}{\partial m}\delta m+\frac{\partial \hat A_{\mu}}{\partial q}\delta q\,.
\end{equation}
\underline{Mass}: In the specific chosen gauge for the $\hat A$ in Eq. \eqref{ex 3 fields}, $\Phi_{_\infty}=0$. By $\eta_{_M}=\{\partial_t,-\Phi_{_\infty}\}$,  the Eq. \eqref{k epsilon tr example 3} yields
\begin{equation}\label{ex 3 M}
\hspace*{0.15cm}\hat \delta M =\frac{1}{8}\times \delta m+ 0\times \delta q  \qquad \Rightarrow \qquad M= \frac{m}{8}\,.
\end{equation}
\underline{Angular momentum}: For $\eta_{_J}=\{-\partial_\varphi,0\}$, 
\begin{equation}
\hat \delta J =0\times \delta m+0\times \delta q  \qquad \Rightarrow \qquad J=0\,.
\end{equation}
\underline{Electric charge}: For the exact symmetry $\eta_{_Q}=\{0,1\}$, 
\begin{equation}\label{ex 3 Q}
\hspace*{0cm}\hat \delta Q=0\times \delta m+ \frac{1}{4G}\times \delta q  \qquad \Rightarrow \qquad Q= \frac{q}{4G}\,.
\end{equation}
\underline{Entropies}: For any horizon present in this geometry, one can associate an entropy. Surface gravities, electric potentials, and  horizon Killing vectors for different horizons would be collectively
\begin{equation}
\kappa_{_\text{H}}=\frac{r_{_\mathrm{H}}}{l^2}-\frac{q^2}{4r_{_\text{H}}}\,,\qquad  \Phi_{_\text{H}}=-\frac{q}{2}\log (\frac{r_{_\mathrm{H}}}{l})\,, \qquad \zeta_{_\text{H}}=\partial_t\,,
\end{equation} 
where $r_{_\mathrm{H}}$ denotes the radius of any one of the horizons. By the choice of $\eta_{_\text{H}}=\frac{2\pi}{\kappa_{_\text{H}}}\{\zeta_{_\mathrm{H}},-\Phi_{_\text{H}}\}$,
\begin{equation}\label{ex 3 S}
\hat \delta S_{_\text{H}}=\frac{2\pi}{\kappa_{_\mathrm{H}}}(\frac{1}{8}\delta m-\frac{\Phi_{_\text{H}}}{4G}\delta q)\,.
\end{equation}
Now, by inserting the relations
\begin{equation}
\frac{\partial r_{_\text{H}}}{\partial m}= \frac{G}{\frac{2r_{_\text{H}}}{l^2}-\frac{q^2}{2r_{_\text{H}}}}\,, \qquad \frac{\partial r_{_\text{H}}}{\partial q}= \frac{q\log (\frac{r_{_\text{H}}}{l})}{\frac{2r_{_\text{H}}}{l^2}-\frac{q^2}{2r_{_\text{H}}}}\,, 
\end{equation}
we find
\begin{equation}
\hat \delta S_{_\text{H}}=\frac{2\pi}{4G}(\frac{\partial r_{_\text{H}}}{\partial m} \delta m +\frac{\partial r_{_\text{H}}}{\partial q} \delta q)=\hat \delta (\frac{2\pi r_{_\text{H}}}{4G}) \qquad \Rightarrow \qquad S_{_\text{H}}=\frac{2\pi r_{_\text{H}}}{4G}\,.
\end{equation}
Reference points for the charges above are chosen to vanish for the pure AdS$_3$ geometry, i.e.  the geometry identified by $m=q=0$. In comparison with the calculations done in the literature (see e.g.  Ref. \cite{Martinez:1999qi}), the charges above are finite, and one does not need to regularize any divergent result. Notice that by replacing $\log(\frac{r}{l})\to \log(\frac{r}{r_0})$ for some $r_0$, the solution would remain a solution, but $M\to M+\frac{q^2}{16 \,G}\log(\frac{l}{r_0})$. \\
\noindent\underline{First laws}: For any one of the horizons, the first law would be
\begin{equation}
\eta_{_\mathrm{H}}=\frac{1}{T_{_\mathrm{H}}}(\eta_{_M}-(\Phi_{_\text{H}}-\Phi_{_\infty})\eta_{_Q})
\quad \xrightarrow{\text{linearity of $\delta H_\epsilon$ in $\epsilon$\,\,\,}}
\quad \delta S_{_\mathrm{H}}=\frac{1}{T_{_\mathrm{H}}}(\delta M-(\Phi_{_\text{H}}-\Phi_{_\infty})\delta Q)\,.
\end{equation}
which can be checked for the parametric variations via Eqs.  \eqref{ex 3 M}, \eqref{ex 3 Q} and \eqref{ex 3 S}.

At the end, it is worth mentioning that in the definition of $\eta_{_\mathrm{H}}$ in the examples above, we tacitly assumed $\kappa_{_\mathrm{H}}\neq 0$. For $\kappa_{_\mathrm{H}}=0$ cases, which are called extremal black holes, one can find infinite number of exact symmetries in their near horizon regions, as generators of the entropy \cite{Hajian:2013lna,HSS:2014twa}. Using any one of these generators, the SPSM would reproduce the entropy for the extremal black holes too. An explicit example for such an analysis can be found in Ref.  \cite{HS:2015xlp}, where the near horizon of the extremal Kerr-Newman black hole is studied.

\section{Conclusion}
In this work, after reviewing the solution phase space method, we applied it to a family of higher curvature gravitational theories. The family which we focused on, contained $f(R)$ gravity, quadratic Riemann and Ricci terms, an arbitrary number of Abelian gauge fields, and arbitrary scalar fields. After elaborating the $\boldsymbol{k}_\epsilon$, which is pragmatically the most important differential form for the calculations, four families of black hole solutions were analyzed. Specifically, their conserved charges were calculated, confirming the results formerly calculated by the other methods. By the way, the results ameliorated the divergence appearing in the calculation of mass for the charged static BTZ black hole. The main advantages of the method are: (1) it works for any higher curvature theory in any dimension, (2) asymptotics and horizons are unimportant in the charge calculations,  (3) conserved charges are automatically regular, (4) conserved charges are unambiguous, (5) all the charges, including the entropy and electric charge, are calculated by a single machinery,  (6) the proof of the first law(s) is very simple.

\paragraph{Acknowledgement:}  K.H would like to thank members of the quantum gravity group at IPM, specifically M.M. Sheikh-Jabbari and M.H. Vahidinia, for useful discussions. He also thanks ICTP for its hospitality during the Workshop on Topics in Three Dimensional Gravity. Some of the calculations in this paper have been done using the HPC cluster in IPM. This work has been supported by the \emph{Allameh Tabatabaii} Prize Grant of \emph{National Elites Foundation} of Iran and the \emph{Saramadan grant} of the Iranian vice presidency in science and technology.

%%%%%%%%%%%%%%%%%%%%%%%%%%%%%%%%%%%%%%%%%%%%%%%%%%%%%%%%%%%%%%%%%%%%%%%%%%%%%%%%%%%%%%%%

\end{document}